\documentclass[times]{article}
\usepackage{amsfonts}
\usepackage{graphicx}
\usepackage{array}
\begin{document}

\pagenumbering{arabic}

\title{Space Efficient Secret Sharing}

\author{Abhishek~Parakh and Subhash Kak\\ \\
Computer~Science~Department, Oklahoma~State~University\\
Stillwater, OK-74078}

\maketitle

\begin{abstract}
This note proposes a method of space efficient secret sharing in which $k$ secrets are mapped into $n$ shares $(n\geq k)$ of the same size. Since, $n$ can be chosen to be equal to $k$, the method is space efficient. This method may be compared with conventional secret sharing schemes that divide a single secret into $n$ shares.
\end{abstract}

\section{Introduction}
A $k$-out-of-$n$ secret sharing scheme conventionally encodes a single secret $s$ in $n$ shares such that any $k$ of them can be used to reconstruct the secret (Shamir's scheme \cite{ref1}). However, Shamir's scheme is not space efficient because it leads to a $n$-fold increase in total storage requirement. Some efforts have been made to create space efficient schemes \cite{ref5, ref6} under certain restrictions, but most other techniques achieve this in the general case by compromising security constraints and adopting a computational security model \cite{ref2, ref3, ref4}, which is weaker compared to that of Shamir's scheme.

We present an algorithm to split $k$ secrets of length $b$ each into $n$ shares such that each share is effectively of size $(n/k)\cdot b$. Since, $n/k$ can be chosen to be close to 1, our algorithm is space efficient.

Our proposed scheme is based on interpolation just like Shamir's scheme, but unlike Shamir we use the $k$ secrets to obtain a polynomial and then create $n$ shares by finding the values of the polynomial at $n$ \textit{new points}. Without knowledge of at least $k$ of these shares, equal to the number of coefficients associated with the polynomial, the original secrets, defined for pre-fixed points of the variable, cannot be found.

\section{Space efficient secret sharing}
The proposed algorithm is as follows. Consider that k secrets $s_0$, $s_1$, ..., $s_{k-1}$ are numbers from the finite field $\mathbb{Z}_p$, where $p$ is prime.\\

\noindent \verb"Algorithm A"
\begin{enumerate}
\item Map $k$ secrets as $k$ distinct points $P_i=(i,s_i)$, $0\leq i\leq (k-1)$.
\item Generate polynomial $f(x)=a_0+a_1x+a_2x^2+...+a_{k-1}x^{k-1}$ of degree $(k-1)$ by interpolating points $P_i$, $0\leq i\leq (k-1)$.
\item Sample the polynomial $f(x)$ at $n$ distinct points $D_i=f(i+k-1)$, where $1\leq i\leq n$.
\item The shares are then given by $(k,D_1)$, $(k+1,D_2)$, ..., $(k+n-1, D_n)$.
\end{enumerate}

Reconstruction of the secrets involves first interpolating any $k$ shares to recreate $f(x)$ and then evaluating it to reveal the secrets $s_i=f(i)$, $0\leq i\leq (k-1)$. All the computations are performed modulo a prime $p>(s_{max},n)$, $s_{max}=max(s_i)$, $0\leq i\leq (k-1)$.

Reconstruction of the secrets in Algorithm A can be viewed as solution to a set of linear equations $A\cdot v=F$, where $A$ is a $k\times k$ Vandermonde matrix \cite{ref7} (generated using the x-coordinates of any $k$ shares), $v$ is a $k\times 1$ vector of unknowns (polynomial coefficients $a_i$'s) and $F$ is the $k\times 1$ vector of the y-coordinates of the shares. Without loss of generality, assume we have $k-1$ shares $(k,D_1)$, $(k+1,D_2)$, ..., $(2k-2,D_{k-1})$. We can explicitly write the system of equations as follows,
\[ \setlength{\extrarowheight}{3pt}
\left[\begin{array}{ccccc}
1& k& k^2& \ldots& k^{k-1}\\
1& k+1& (k+1)^2& \ldots& (k+1)^{k-1}\\
\vdots& \vdots& \vdots& \vdots& \vdots\\
1& 2k-2& (2k-2)^2& \ldots& (2k-2)^{k-1} \end{array}
\right]\cdot
\left[\begin{array}{c}
a_0\\
a_1\\
\vdots\\
a_{k-1} \end{array}
\right]=
\left[\begin{array}{c}
D_1\\
D_2\\
\vdots\\
D_{k-1} \end{array}
\right]
\]

Clearly to find the value of $a_i$'s one would need to invert the Vandermonde matrix (find $A^{-1}$) and also know the values of $D_i$, $1\leq i\leq k-1$. Further, since we have distributed the secrets among the coefficients of the polynomials, one needs to reconstruct the coefficients $a_i$, $0\leq i\leq (k-1)$, in order to obtain any and all of the secrets.

Figure 1 illustrates that $p$ polynomials remain equally probable if we have the knowledge of only 3 shares (out of 4 required) shares (points).

\pagebreak
\begin{figure}[htb!]
\centering%
\includegraphics[scale=.47]{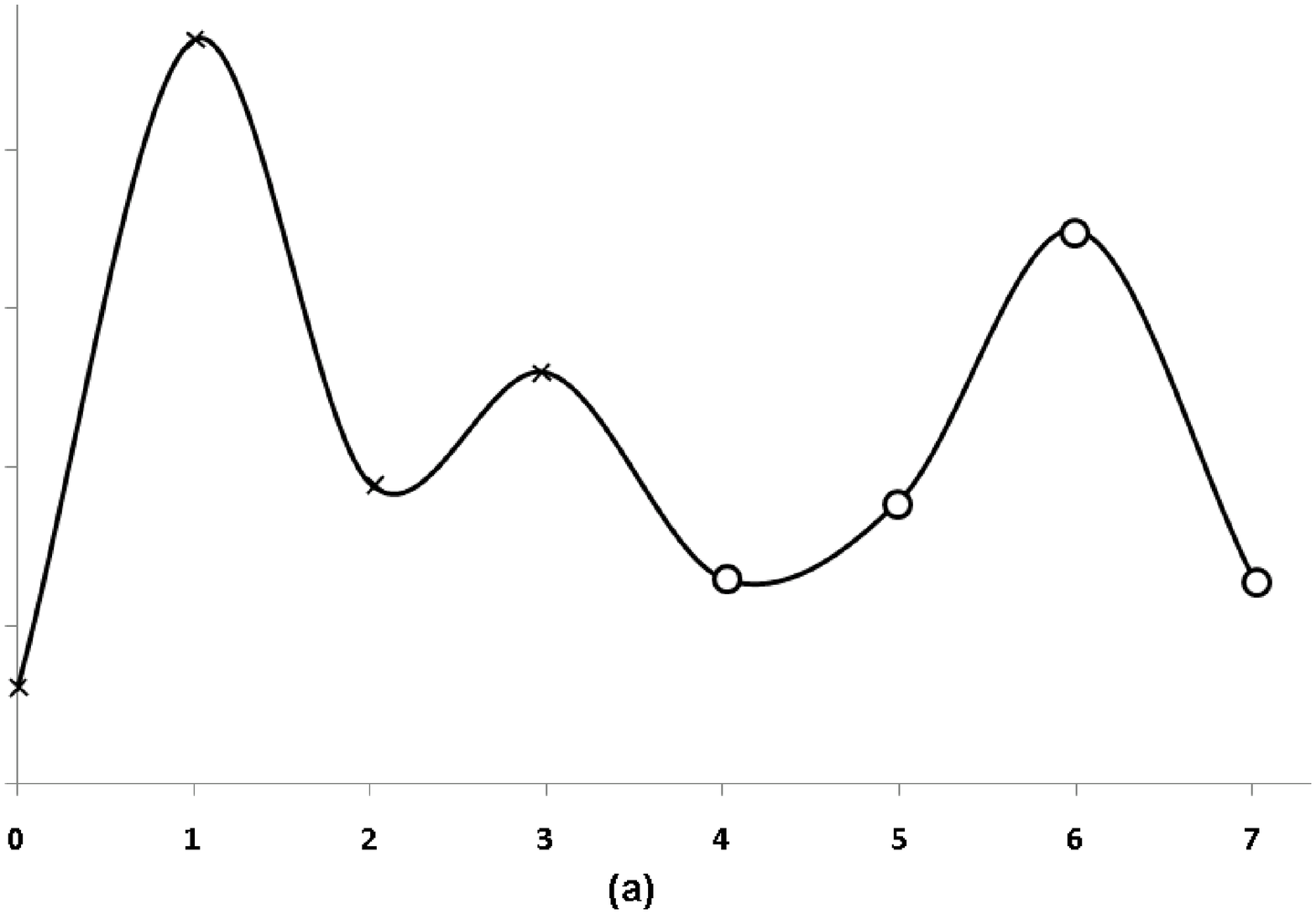}
\label{fig:graph1}
\end{figure}

\begin{figure}[htb!]
\centering%
\includegraphics[scale=.50]{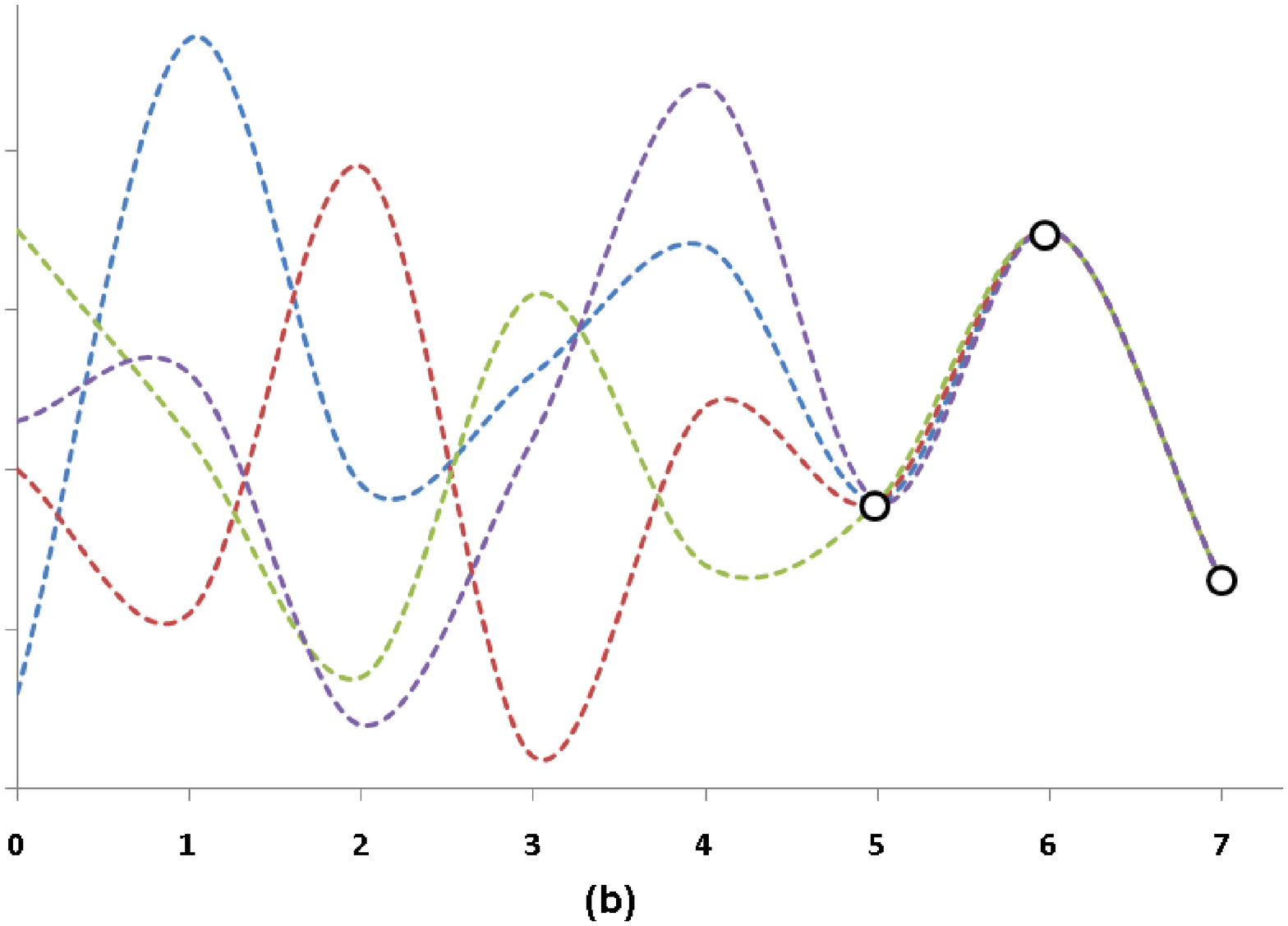}
\caption{(a) x: Four secrets; o: Four shares. (b) Three shares, one fewer than the required number, are known, but they leave $p$ polynomials equally likely.}
\label{fig:graph1}
\end{figure}
\pagebreak

\noindent \textbf{Example.}
Let $s_0=10$, $s_1=23$, $s_2=16$, and $s_3=25$ be four secrets that are to be shared between 6 players such that any 4 of them can reconstruct all the 4 secrets, i.e. $n=6$ and $k=4$. Let $p=31$. We apply Algorithm 2 as follows,

\begin{enumerate}
\item Map the 4 secrets as 4 distinct points $P_1=(0,10)$, $P_2=(1,23)$, $P_3=(2,16)$, and $P_4=(3,25)$.
\item Generate a polynomial $f(x)$ by interpolating points $P_1$, $P_2$, $P_3$, and $P_4$ as follows,
\noindent\[\setlength{\extrarowheight}{10pt}
\begin{array}{rl}
f(x)&=\displaystyle\sum_{i=0}^3 s_i \displaystyle\prod_{j=0,j\neq i}^3 \frac{x-x_j}{x_i-x_j} \hspace{7mm} (mod \hspace{1.5mm}31)\\
&=10\cdot \frac{(x-1)(x-2)(x-3)}{(0-1)(0-2)(0-3)}+23\cdot \frac{(x-0)(x-2)(x-3)}{(1-0)(1-2)(1-3)}\\ &\hspace{15mm}+16\cdot \frac{(x-0)(x-1)(x-3)}{(2-0)(2-1)(2-3)}+25\cdot \frac{(x-0)(x-1)(x-2)}{(3-0)(3-1)(3-2)}\\
&=(19x^3-21x^2+23x-21)+(27x^3-11x+7x)\\
&\hspace{15mm}+(23x^3-30x^2+7x)+(30x^3-28x^2+29x)\\
&=6x^3+3x^2+4x-21 \hspace{7mm}(mod \hspace{1.5mm}31)
\end{array}\]
\item Sample $f(x)$ at 6 distinct points $D_1=f(4)=24$, $D_2=f(5)=18$, $D_3=f(6)=12$, $D_4=f(7)=11$, $D_5=f(8)=20$, and $D_6=f(9)=13$.
\item The shares are given by (4,24), (5,18), (6,12), (7,11), (8,20), and (9,13).
\end{enumerate}

This completes the construction of the shares by the algorithm.

The reconstruction phase uses any four shares from step 4 above and interpolates them to compute $f(x)$. Then sampling $f(x)$ at points $x$=0, 1, 2 and 3 would reveal $s_0$, $s_1$, $s_2$, and $s_3$ respectively.

If an adversary or colluding players determine three shares out of the required four shares in the above example, only a quadratic polynomial can be reconstructed leaving 31 possibilities for the fourth share equally likely. Each of these 31 possibilities give rise to 31 unique cubic polynomials. Only one of these 31 polynomials will pass through all the secrets. Hence the probability of computing the secrets even after the knowledge of $k-1=3$ shares is $\frac{1}{p}=\frac{1}{31}$.

In practice, the primes chosen are very large (on the order of 1024 bits). Since the secrets can be random numbers from the field, the knowledge of $k-1$ shares leaves $p$ possibilities for the last share, which can only be guessed correctly with a small probability on the order of $\frac{1}{2^{1024}}$. Consequently, knowledge of only $k-2$ shares reduces the probability to $(\frac{1}{2^{1024}})^{2}$, and so on in general if $t$ shares are known then the secret can be guessed with a small probability of $(\frac{1}{2^{1024}})^{k-t}$.

This level of security may be sufficient in many applications, such as the sensor networks where at any given time only a few sensor may be compromised.

A variant of Algorithm A is the case where the interpolating points are randomly chosen with the constraint that they don't overlap with the x-coordinates of the shares.

\section{Conclusions}
We have presented a space efficient secret sharing scheme. The proposed method can also be used to encode a large secret, by dividing the secret into smaller pieces that are shared between several parties. Further it may be used for secure transmission of data over parallel communication channels or secure online data storage, providing efficiency higher than the method of Garay et al. \cite{ref8}.


\begin{thebibliography}{7}
\bibitem{ref1}
A. Shamir. How to share a secret. Communications of ACM, vol. 22, issue 11, pages 612-613, 1979.

\bibitem{ref2}
Bruce Schneier. Schneier's Cryptography Classics Library: Applied Cryptography, Secrets and Lies, and Practical Cryptography. Wiley, 2007.

\bibitem{ref3}
Philip Rogaway and Mihir Bellare. Robust computational secret sharing and a unified account of classical secret-sharing goals. Proceedings of the 14th ACM conference on Computer and Communications Security. Pages 172-184, 2007.

\bibitem{ref4}
V. Vinod et al. On the power of computational secret sharing. Indocrypt 2003, vol. 2904, pages 265-293, 2003.

\bibitem{ref5}
M. Gnanaguruparan and S. Kak. Recursive hiding of secrets in visual cryptography. Cryptologia, vol. 26, pages 68-76, 2002.

\bibitem{ref6}
A. Parakh and S. Kak. A recursive threshold visual cryptography scheme. Cryptology ePrint Archive, Report 535, 2008.

\bibitem{ref7}
L. R. Turner. Inverse of the Vandermonde Matrix with Applications. Glenn Research Center, NASA, 1966.

\bibitem{ref8}
Juan A. Garay, Rosario Gennaro, Charanjit Jutla and Tal Rabin. Secure distributed storage and retrieval. Theoretical Computer Science. Pages 275-289, 1997.

\end{thebibliography}
\end{document}